\documentclass[pre,aps,twocolumn,superscriptaddress,floats,floatfix,10pt]{revtex4-1}

\usepackage[english]{babel}
\usepackage{amsmath,amsfonts,amssymb}
\usepackage{graphicx}  

\usepackage{dcolumn}
\usepackage{bm}
\usepackage{textcomp}
\usepackage{color}
\usepackage{float}

\begin{document}

\title{Plasma Equilibrium inside Various Cross-Section Capillary Discharges}

\author{G.\,Bagdasarov}
\affiliation{Keldysh Institute of Applied Mathematics RAS, Moscow, 125047, Russia}
\affiliation{National Research Nuclear University MEPhI (Moscow Engineering Physics Institute), Moscow, 115409, Russia}

\author{P.\,Sasorov}
\affiliation{Keldysh Institute of Applied Mathematics RAS, Moscow, 125047, Russia}

\author{A.\,Boldarev}
\affiliation{Keldysh Institute of Applied Mathematics RAS, Moscow, 125047, Russia}
\affiliation{National Research Nuclear University MEPhI (Moscow Engineering Physics Institute), Moscow, 115409, Russia}

\author{O.\,Olkhovskaya}
\affiliation{Keldysh Institute of Applied Mathematics RAS, Moscow, 125047, Russia}

\author{V.\,Gasilov}
\affiliation{Keldysh Institute of Applied Mathematics RAS, Moscow, 125047, Russia}
\affiliation{National Research Nuclear University MEPhI (Moscow Engineering Physics Institute), Moscow, 115409, Russia}

\author{A.\,J.\,Gonsalves}
\affiliation{Lawrence Berkeley National Laboratory, Berkeley, California 94720, USA}
\author{S.\,Barber}
\affiliation{Lawrence Berkeley National Laboratory, Berkeley, California 94720, USA}
\author{S.\,S.\,Bulanov}
\affiliation{Lawrence Berkeley National Laboratory, Berkeley, California 94720, USA}
\author{C.\,B.\,Schroeder}
\affiliation{Lawrence Berkeley National Laboratory, Berkeley, California 94720, USA}
\author{J.\,van\,Tilborg}
\affiliation{Lawrence Berkeley National Laboratory, Berkeley, California 94720, USA}
\author{E.\,Esarey}
\affiliation{Lawrence Berkeley National Laboratory, Berkeley, California 94720, USA}
\author{W.\,P.\,Leemans}
\affiliation{Lawrence Berkeley National Laboratory, Berkeley, California 94720, USA}

\author{T.\,Levato}
\affiliation{Institute of Physics ASCR, v.v.i. (FZU), ELI-Beamlines Project, 182 21 Prague, Czech Republic}
\author{D.\,Margarone}
\affiliation{Institute of Physics ASCR, v.v.i. (FZU), ELI-Beamlines Project, 182 21 Prague, Czech Republic}
\author{G.\,Korn}
\affiliation{Institute of Physics ASCR, v.v.i. (FZU), ELI-Beamlines Project, 182 21 Prague, Czech Republic}

\author{S.\,V.\,Bulanov}
\affiliation{National Institutes for Quantum and Radiological Science and Technology (QST), Kansai Photon Science Institute, 8-1-7 Umemidai, Kizugawa, Kyoto 619-0215, Japan}

\begin{abstract}
Plasma properties inside a hydrogen-filled capillary discharge waveguide were modeled with dissipative magnetohydrodynamic simulations to enable analysis of capillaries of circular and square cross-sections implying that square capillaries can be used to guide circularly-symmetric laser beams. When the quasistationary stage of the discharge is reached, the plasma and temperature in the vicinity of the capillary axis has almost the same profile for both the circular and square capillaries. The effect of cross-section on the electron beam focusing properties were studied using the simulation-derived magnetic field map. Particle tracking simulations showed only slight effects on the electron beam symmetry in the horizontal and diagonal directions for square capillary
\end{abstract}

% add showkeys and/or showpacs to include the next blocks into the paper
 % Laser-plasma acceleration of electrons and ions

\maketitle

\section{Introduction\label{sec:intro}}

For several decades capillary discharges have been under intensive investigation due to various promising applications, such as waveguides for laser electron accelerators and X-ray lasers~\cite{Ehrlich1996,ECL2009,Leemans2014,Benware1998,Steinke2016}, and more recently for focusing of electron beams~\cite{Tilborg2015,Pompili2017,Tilborg2017}. The majority of the experiments use circular cross-section capillaries, which reduce the dimensionality of the problem under consideration, simplifying the theoretical and computer simulation studies, and allowing the use of 1D MHD codes~\cite{Tilborg2015,Tilborg2017,Bobrova2001,Broks2005,Bobrova2013,Gonsalves2016}. On the other hand, square cross-section capillaries, which have attracted substantially less attention, have several advantages for transverse plasma diagnostics~\cite{Gonsalves2007}.

For optical waveguides, the transverse electron density profile determines the guiding properties. A useful parameter to describe the guiding property is the matched spot size of the channel, which in the low power limit is the beam size that will propagate without change in transverse dimension. For a Gaussian beam and a parabolic channel, the matched spot size is given by~\cite{Bobrova2001}:
\begin{displaymath}
W_m=(0.5 \pi r_e \partial^2 n_e /\partial r^2)^{-1/4},
\end{displaymath}
where $n_e$ is the electron density, and $r_e = e^2/m_ec^2 \approx 2.817\cdot10^{-13}$\,cm is the classical electron radius. The guiding density profile inside a hydrogen-filled capillary discharge waveguide was first simulated using 1D MHD simulations~\cite{Bobrova2001,Broks2005}. Subsequently 2D simulations for capillaries of square cross-section~\cite{Broks2007} showed that for a capillary with transverse size 0.465\,mm the matched spot size of the channel was 0.065\,mm in the direction perpendicular to the capillary wall, and 0.066\,mm in the diagonal of the square capillary. This result suggested that for this size of capillary, a square cross-section is suitable for guiding circularly-symmetric laser pulses. Furthermore it was found that for circular capillaries with diameter equal to the square capillary width, there was negligible change to the matched spot size. For capillaries of smaller transverse size, one would expect the cross-section to have a larger effect on the guiding properties.

In the present paper, square capillaries of width $500\,\mu$m and $250\,\mu$m are investigated. In addition the effect of cross-section shape on electron beam focusing properties is analyzed via the MHD-derived magnetic field profile, which determines the focusing strength and quality. One of the aims of our paper is to compare the plasma density and temperature distribution formed at the quasistationary stage of the discharge inside the hydrogen-filled capillaries with circular and square cross-sections under almost the same conditions characterizing the initial configurations and the external electric circuit.

\section{Simulation configuration and parameters\label{sec:config}}

We use MHD code MARPLE~\cite{MARPLE12} for simulations of the capillary discharges with various cross-sections. The physical model implemented in the code is similar to the model formulated in~\cite{Bobrova2001}. The plasma is described within the framework of one-fluid, two-temperature (ion and electron) magnetohydrodynamics taking into account the dissipative processes as it follows. The model implements the electron and ion thermal conductivity, the electron-ion energy exchange, the magnetic field diffusion due to final electric conductivity, the Joule heating, and the radiation losses. The equation of state and dissipative coefficients incorporate the degree of the gas ionization.

Since the magnetic field lines penetrate through the interface between the discharge plasma and the insulator, it is required to calculate the magnetic field also inside the insulator domain. Due to this reason the simulation region comprises of two domains, namely, the internal domain filled by plasma (called the ``plasma'') and the external domain called the ``insulator''. To calculate the magnetic field evolution we solve the same MHD equations in the ``plasma'' and ``insulator'' regions. In contrast to the ``plasma'' domain, in the ``insulator'' region the medium is treated as immobile (${\bf v} = 0$) and with zero conductivity ($\sigma = 0$) fluid.

\begin{figure}[h!t]
  \includegraphics[width=\linewidth]{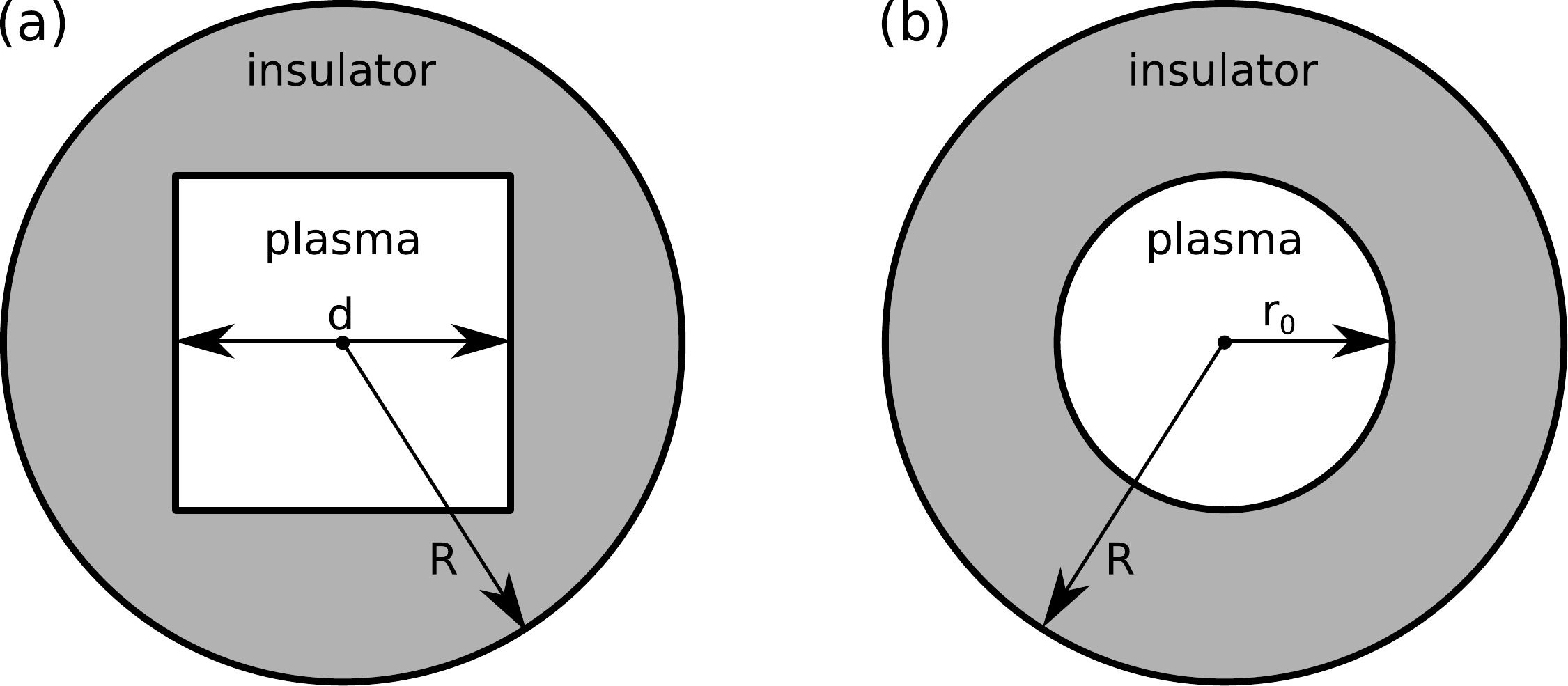}
  \caption{Computational domains for circular (a) and square (b) capillaries. The internal ``plasma'' domains have the transverse size equal to $d = 2r_0$. The external ``insulator'' regions are of the radius $R$.\label{fig:sketch}}
\end{figure}

Two cross-section shapes of the capillaries were considered in our simulations: circular and square shapes, as shown in Fig.\,\ref{fig:sketch}.

In the simulations we consider the capillaries with the diameter/size equal to $d = 2r_0 = 500$ and $250\,\mu$m, prefilled with pure hydrogen plasma of homogeneous initial (at $t=0$) density equal to $\rho_0 = 3.5\cdot10^{-6}\,\mathrm{g/cm^3}$. Initially there is no current inside the channel and hence, in order to initiate the discharge, the hydrogen is assumed to be slightly ionized ($T_e = T_i = 0.5$\,eV). The same temperature was set on the wall. The external passive insulator domain is of the radius equal to $R = 500\,\mu$m. The discharges are driven by the same pulse of the electric current with approximately sinusoidal profile having the quarter-period of 180\,ns and the peak current of 311\,A. The simulation time is within the range, $t\in\left[0, 250\right]$\,ns. The chosen capillary, plasma and external electric circuit parameters correspond to those discharge parameters that are typical for experiments on laser wake-field electron acceleration for guiding ultrashort high power laser pulses (see Ref.\,\cite{Leemans2014} and literature cited therein).

\section{Simulation results\label{sec:results}}

\begin{figure}[h!t]
  \includegraphics[width=0.9\linewidth]{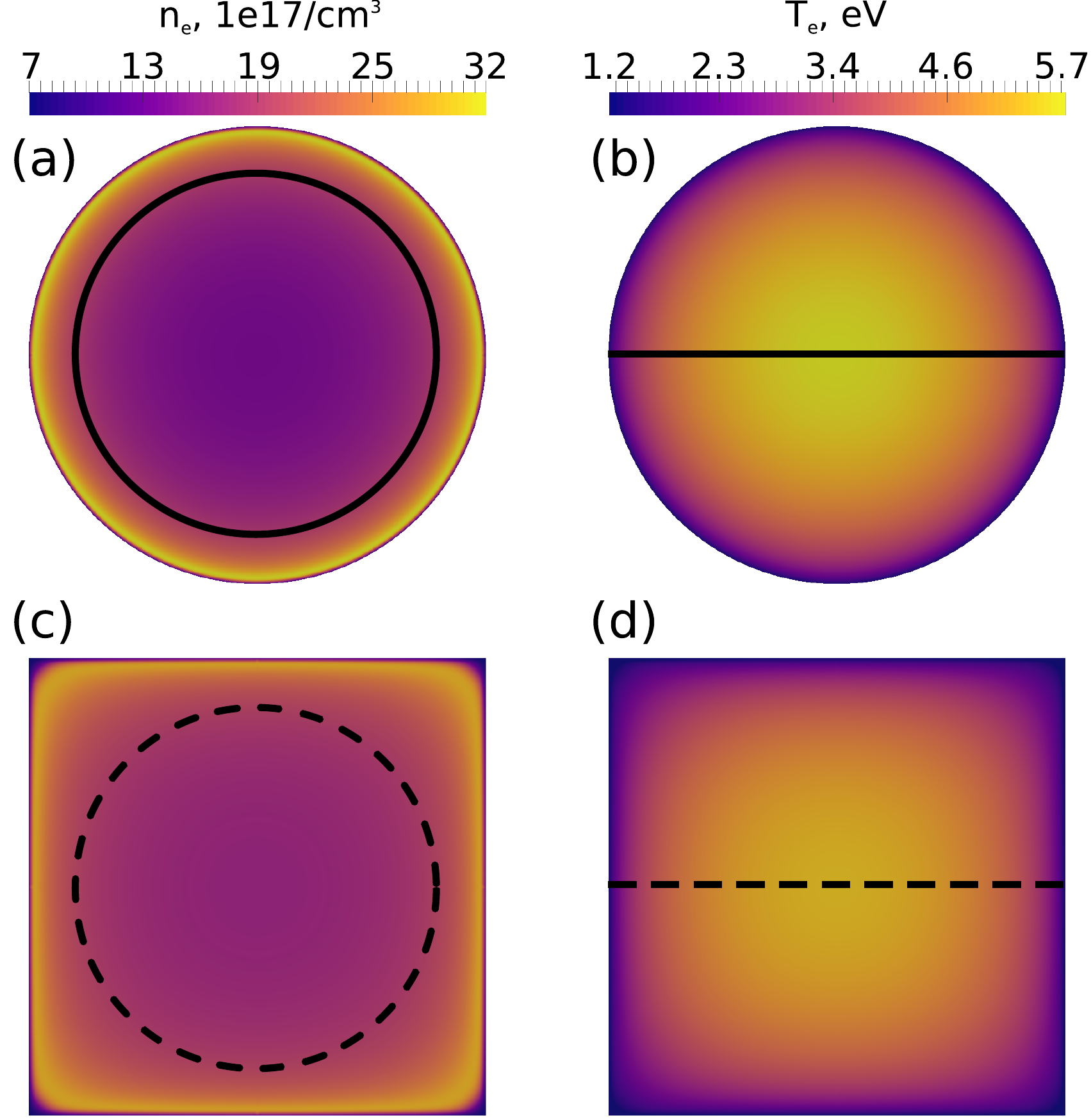}
  \caption{Distribution of the electron density and temperature at $t=200$\,ns in the $(x,y)$ plane. For the circular cross-section capillary $T_e$ and $n_e$ are shown in frames (a) and (b), for the square cross-section capillary~-- in frames (c) and (d).\label{fig:nT2D}}
\end{figure}

Figure\,\ref{fig:nT2D} presents the electron density and temperature distributions for both circular and square capillaries at the same time $t = 200$\,ns (20\,ns after current has reached its maximum), when the quasistatic equilibrium stage of the discharge is reached. The results obtained for the circular cross-section capillary are in a good agreement with the experiments and simulations conducted previously (see~\cite{Tilborg2015,Tilborg2017,Bobrova2001,Broks2005,Bobrova2013,Gonsalves2016}).

\begin{figure}[h!t]
  \includegraphics[width=\linewidth]{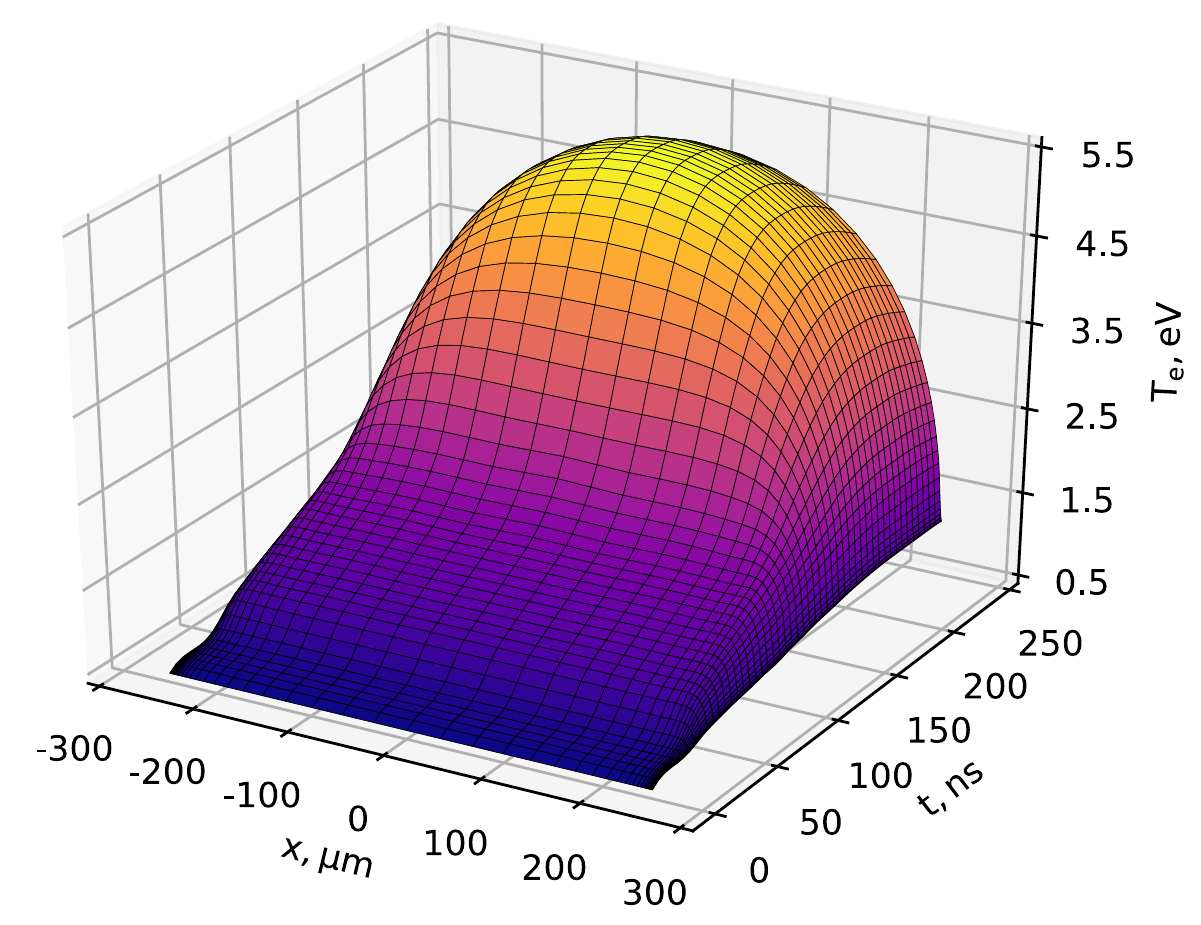}
  \caption{Time evolution of the electron temperature inside the square cross-section capillary at the dash line in Fig.\,\ref{fig:nT2D}\,(d).\label{fig:Te-t}}
\end{figure}

\begin{figure}[h!t]
  \includegraphics[width=\linewidth]{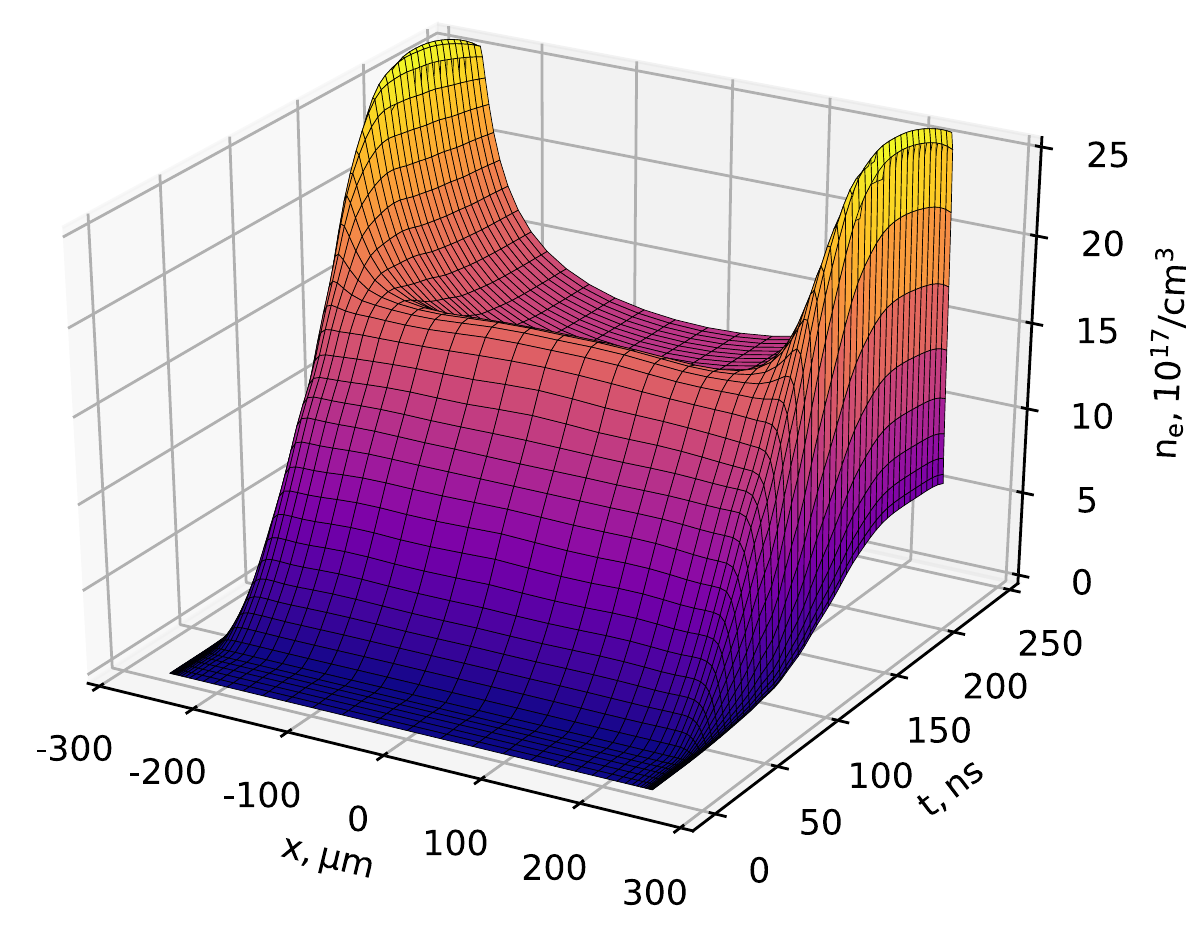}
  \caption{Time evolution of the electron density inside the square cross-section capillary at the dash line in Fig.\,\ref{fig:nT2D}\,(d).\label{fig:ne-t}}
\end{figure}

The time evolution of the electron density and temperature (taken at the dash line shown in Fig.\,\ref{fig:nT2D}) in the square cross-section capillary is presented in Fig.\,\ref{fig:Te-t} and Fig.\,\ref{fig:ne-t}, respectively. At about 200\,ns a near parabolic temperature profile with the temperature maximum at the capillary axis is formed (Fig.\,\ref{fig:Te-t}). The corresponding electron density distribution has a near-parabolic profile with the density minimum at the capillary axis (Fig.\,\ref{fig:ne-t}).

\begin{figure}[h!t]
  \includegraphics[width=0.9\linewidth]{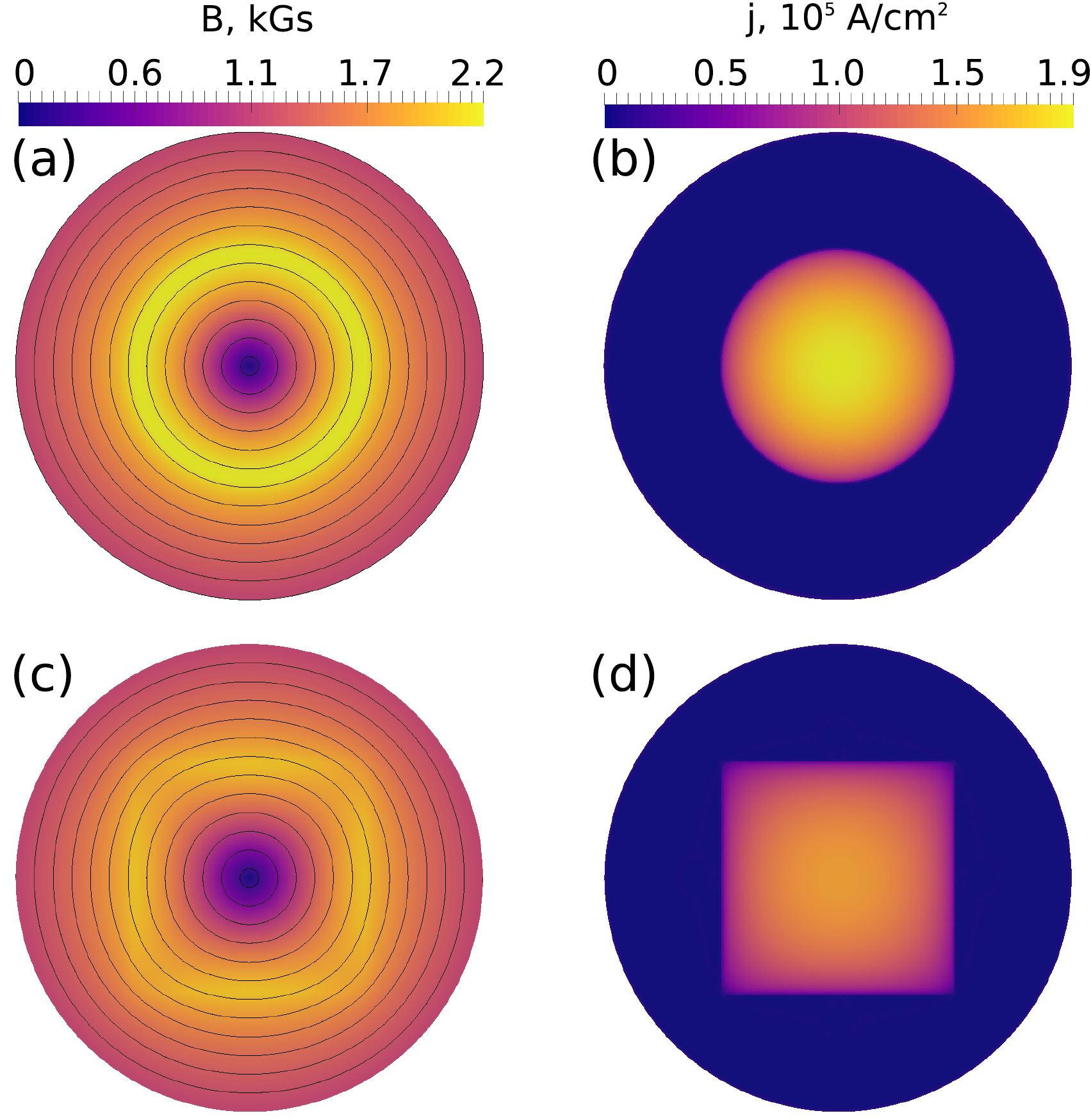}
  \caption{Magnetic field strength and magnetic field lines (left figures) and current density distributions (right figures) inside the capillaries at $t=200$\,ns. Frames (a) and (b) correspond to the circular cross-section capillary, and frames (c) and (d) show the magnetic field and current density distribution inside the square cross-section capillary.\label{fig:Bj}}
\end{figure}

To compare the circular and square cross section capillaries, in Fig.\,\ref{fig:Bj} we show the distributions of magnetic field and current density inside the capillaries at 200\,ns. Frames (a) and (b) in Fig.\,\ref{fig:Bj} correspond to the circular cross-section capillary, and frames (c) and (d) in the same figure show the magnetic field and current density distribution, respectively, inside square cross-section capillary.

As seen in Fig.\,\ref{fig:Bj}\,(c), the magnetic field lines are not perfectly circular near the plasma-insulator interface for the square capillary. For the circular one the field lines are perfectly circular, as expected. We note that near the axis ($r\to 0$) of the square capillary and near the external boundary of the insulator ($r\approx R$) the magnetic field lines become almost circular. These indicate that the size of insulator region was chosen correctly.

\begin{figure}[h!t]
  \includegraphics[width=\linewidth]{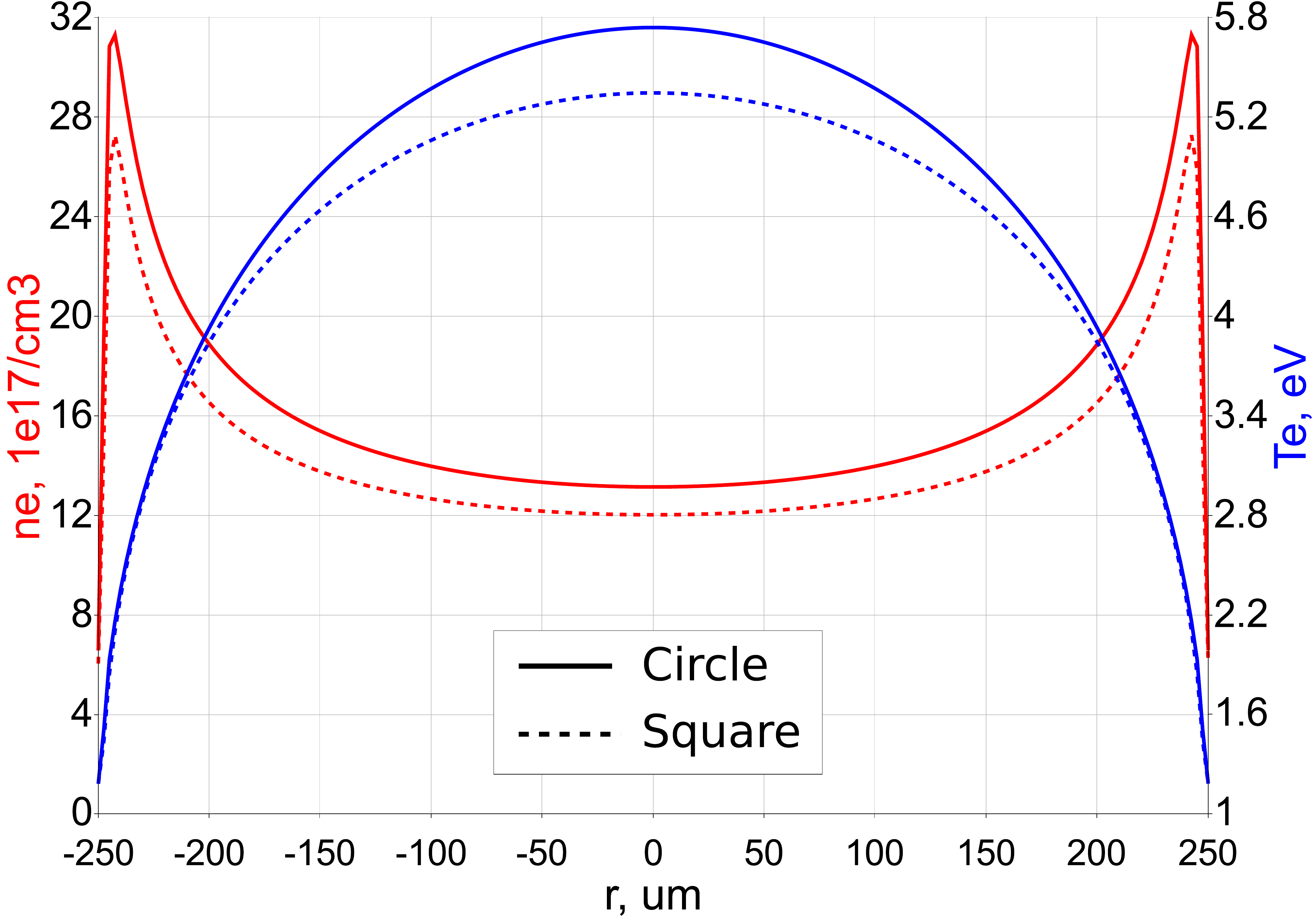}
  \caption{Electron density and temperature at $t=200$\,ns along the line $y=0$ (they are shown by solid and dash lines in Fig.\,\ref{fig:nT2D} (b) and (d), respectively).\label{fig:nT1D}}
\end{figure}

\begin{figure}[h!t]
  \includegraphics[width=\linewidth]{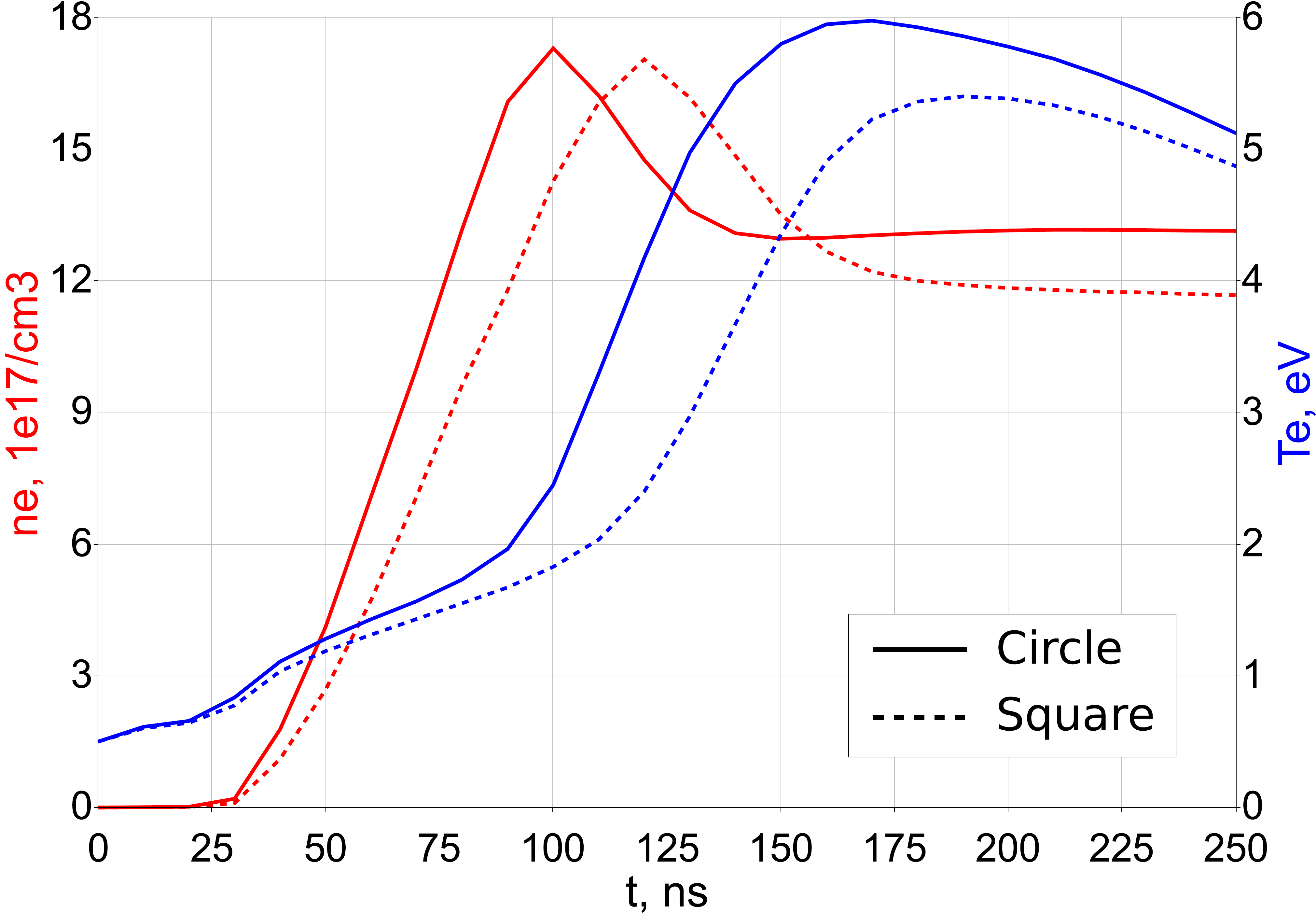}
  \caption{Time-dependencies of electron density and temperature at the capillary axes.\label{fig:nT0-t}}
\end{figure}

In the case of the square cross-section capillary, the values of the temperature ($T_e = 5.34$\,eV) and density ($n_e = 12.0\cdot10^{17}\,\mathrm{cm^{-3}}$) at the capillary axis are lower than for the circular one ($T_e = 5.73$\,eV and $n_e = 13.1\cdot10^{17}\,\mathrm{cm^{-3}}$) as seen in Fig.\,\ref{fig:nT1D}. These differences are caused by differences in the geometry, in particular the electron temperature is lower because current density for the square capillary is less than for the circular one. However, general profiles of plasma and time-dependence of axial plasma parameters for the square capillary is quite similar to that of the circular one (see Fig.\,\ref{fig:nT0-t}).

Using these simulation results, we can approximate the electron density distribution in the vicinity ($r \leq 150\,\mu$m) of the capillary axis for the circular cross-section ($\bigcirc$) and square cross-section ($\square$) capillaries in the same way as it was done in Ref.\,\cite{Bobrova2001} for a circular capillary. They read:

\begin{equation}\label{eq:nesq}
\begin{aligned}
n_e^\bigcirc(r) = n_e(0)&\bigg[1 + 0.33\left(\frac{r}{r_0}\right)^2 + 0.4\left(\frac{r}{r_0}\right)^4 + \dots\bigg],
\\
n_e^\square(r) = n_e(0)&\bigg[1 + 0.29\left(\frac{r}{r_0}\right)^2 +
\\
&{} + \left(0.24 + 0.075\cos 4\varphi\right)\left(\frac{r}{r_0}\right)^4 + \dots\bigg],
\end{aligned}
\end{equation}
where $\varphi$ is the azimuthal angle. The value of coefficient in front of $(r/r_0)^2$~-- 0.33~-- for the circular capillary is coincide with the value obtained in Ref.\,\cite{Bobrova2001}. We note here that the $r^4$ dependence of the transverse density distribution can in principle be detected by observing the laser centroid oscillations as it was proposed in Ref.\,\cite{Gonsalves2010}.

\begin{table}[h!] \centering
\begin{tabular}{ c c c c c }
d($\mu$m) & Shape & $W_m$ ($\mu$m) & $W_m$ diagonal ($\mu$m) & Q.S. scaling \\ 
500 & Circular & 60.3 & 60.3 & 61.5\\  
500 & Square & 65.1 & 65.5 & N.A. \\ 
250 & Circular & 41.9 & 41.8 & 43.5 \\  
250 & Square & 44.8 & 45.8 & N.A.
\end{tabular}
\caption{Matched spot size, $W_m$ as a function of capillary diameter $d$, shape, and angle. Here Q.S. scaling stands for the result obtained from the analytical model from Ref.\,\cite{Bobrova2001}.}
\label{Table1}
\end{table}

\begin{figure}[h!t]
  \includegraphics[width=\linewidth]{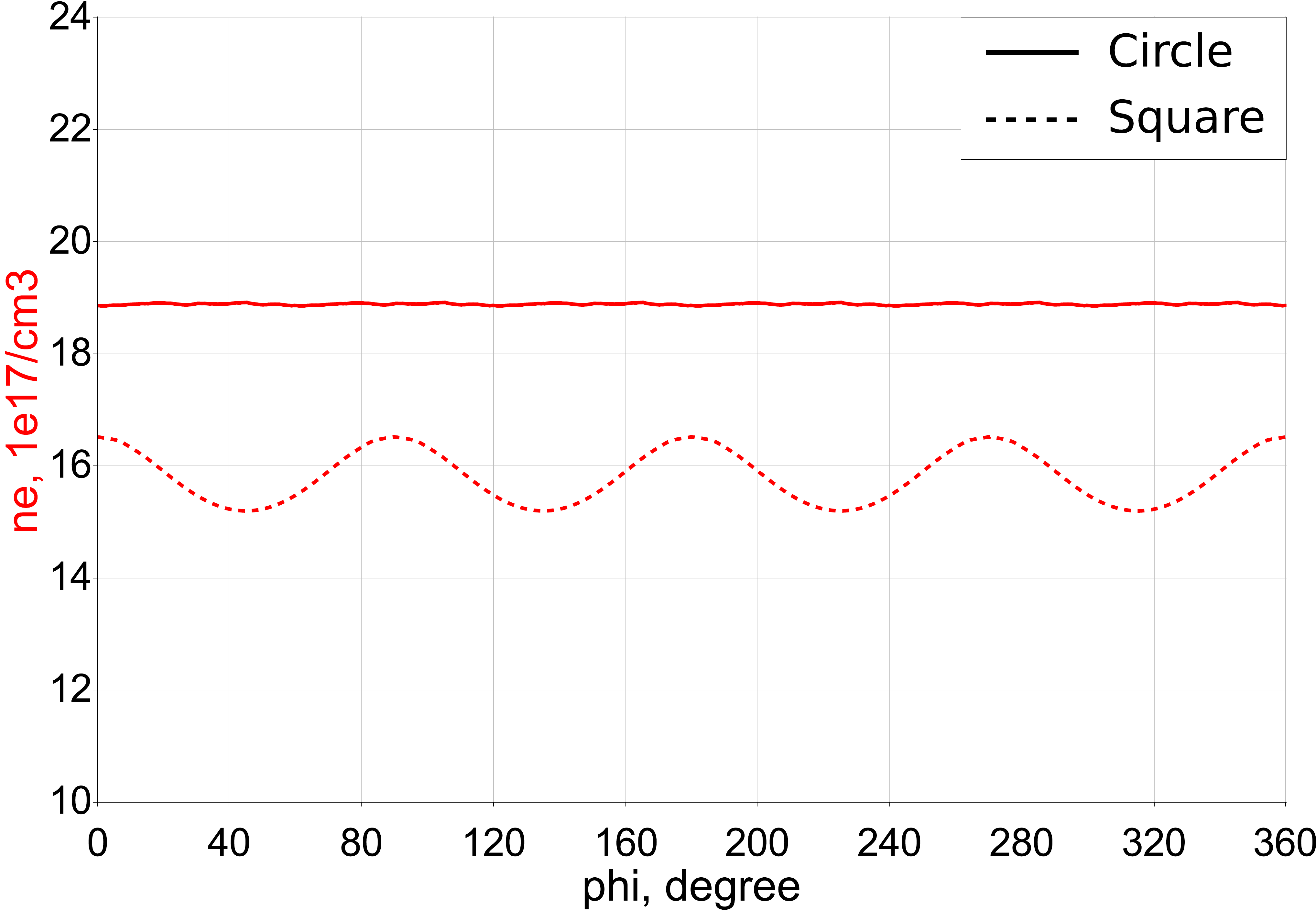}
  \caption{Electron density along circumferences of the same radius $r = 0.02$\,cm (shown by solid and dash lines in Fig.\,\ref{fig:nT2D} (a) and (c), respectively).\label{fig:nphi}}
\end{figure}

The azimuthal inhomogeneity of the electron density near the axis of the square cross-section capillary is substantially weak, as indicated by the smallness of the coefficient in front of the $\cos 4\varphi$ term in Eq.\,\eqref{eq:nesq}. However the further from the axis, the more noticeable the effect of the cosine term becomes (see Fig.~\ref{fig:nphi}). To quantify such effect in the case of laser guiding, the matched spot size calculated according to Ref.\,\cite{Benedetti2012} is shown in table~\ref{Table1} for the different capillaries in the horizontal and diagonal directions, as well as the matched spot size obtained from the analytical quasistationary model~\cite{Bobrova2001}:
\begin{equation}
W_{Q.S.}=1.48\times 10^5\frac{\sqrt{r_0(\mu m)}}{(n_0 [cm^{-3}])^{1/4}}.
\end{equation} 
For both the larger and smaller capillaries, the difference between the matched spot size calculated in the two directions is small, suggesting that even at diameter 250\,$\mu$m, a square capillary is an effective waveguide for circularly-symmetric beams.

\begin{figure}[h!t]
\includegraphics[width=\linewidth]{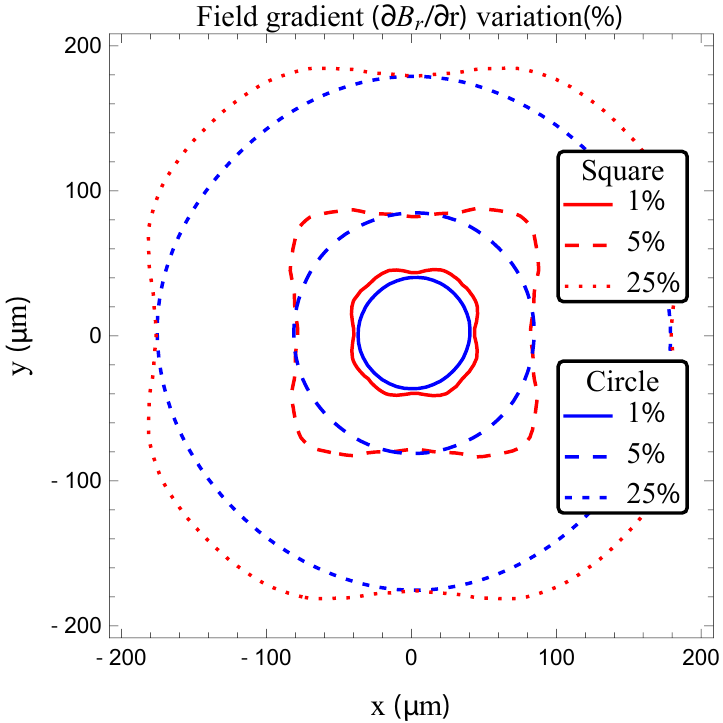}
  \caption{Magnetic field gradient variation from the on-axis value inside the capillaries at $t=200$\,ns, when the quasistatic plasma equilibrium configuration is reached.\label{Bgrad}}
\end{figure}

\begin{figure}[h!t]
\includegraphics[width=0.95\linewidth]{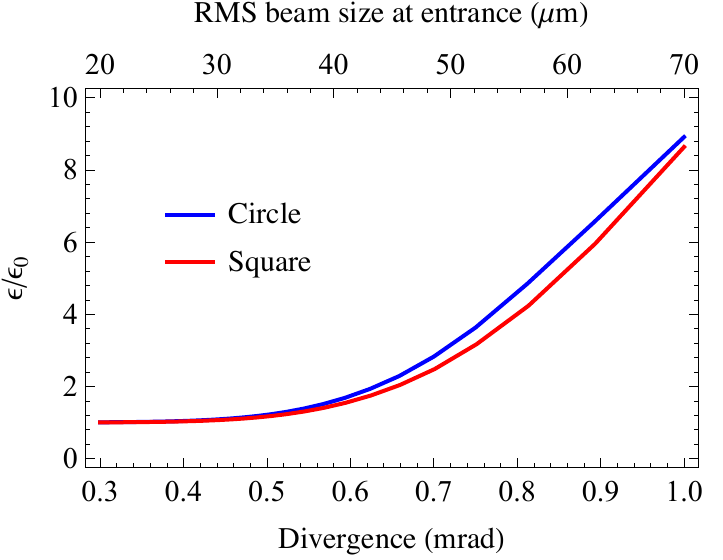}
  \caption{Emittance growth as a function of input electron beam divergence for the capillaries shown in Fig.\,\ref{Bgrad}.\label{emittance}}
\end{figure}

\smallskip

For the electron beam focusing application of discharge capillaries, the focal length of the lens varies inversely with the magnetic field gradient and capillary length. For an ideal lens, the field gradient (and hence focal length) would be constant. For the analysis a factor of beam focusing 0.81 was applied to the magnetic field inside the circular capillary to keep the focal lengths the same and facilitate comparison. Fig.\,\ref{Bgrad} shows the field gradient variation from on-axis value as a function of position for the parameters of Fig.\,\ref{fig:nT2D}. One can see that the variations are similar for the circular and square capillaries, with the square capillary showing a lower variation along the diagonal.

The effect on electron beam focusing of the differences in magnetic field gradient were studied using ELEGANT simulations~\cite{ELEGANT}. Electron beam parameters relevant to the LPA-based XUV FEL under construction at the BELLA center at LBNL were chosen~\cite{XFEL BELLA}. The electron beam energy and source size were 300\,MeV and 0.5~micron, respectively. The electron beam was taken to have no energy spread and the distance from the electron beam waist to the 15\,mm-long capillary was 7\,cm, providing a focal point approximately 1\,m from the capillary exit. The emittance growth due to the field gradient variations of the plasma lens are shown in Fig.\,\ref{emittance}, and as suggested by the field gradient variation, the difference between the circular and square capillaries is small. In both cases for a sufficiently small beam at the entrance of the capillary (r.m.s beam size $\lesssim$10\% of the capillary radius), the emittance growth is negligible since the field gradient variation is small. As the beam fills a greater portion of the capillary, the emittance grows by approximately a factor of 3 when the r.m.s beam size is~$\approx 20$\% of the capillary radius. The square capillary shows a slightly lower emittance growth due to lower field gradient variation along the diagonal. For higher divergence, the difference becomes less due to increased charge loss (percent level) at the entrance of the circular capillary.

\section{Conclusions\label{sec:outro}}

In conclusion, we have simulated  the development of a hydrogen-filled capillary discharge with circular and square cross-sections under the same initial plasma density, capillary size, and the parameters of the external electric circuit. We have found that the calculated magnetic field, electron temperature, and density distribution in the near-axis region of the square and circular capillaries are similar. These results indicate that square capillaries, which allow greater diagnostic accessibility, can be employed to guide cylindrically symmetric laser pulses and focus electron beams.

\section*{ACKNOWLEDGMENTS}

The work was supported in part by the Russian Foundation for Basic Research (Grant No.\,15-01-06195), by the Competitiveness Program of National Research Nuclear University MEPhI (Moscow Engineering Physics Institute), contract with the Ministry of Education and Science of the Russian Federation No.\,02.A03.21.0005, 27.08.2013 and the basic research program of Russian Ac. Sci. Mathematical Branch, project 3-OMN RAS. The work at Lawrence Berkeley National Laboratory was supported by US DOE under contract No.\,DE-AC02-05CH11231. This research was also sponsored by the project ELI~-- Extreme Light Infrastructure–Phase\,2 (CZ.02.1.01/0.0/0.0/15\_008/0000162) through the European Regional Development Fund and by the Ministry of Education, Youth and Sports of the Czech Republic (National Program of Sustainability II Project No. LQ1606).

\end{document}